\documentclass[fp,twocolumn]{jpsj3}

\usepackage{color}
\usepackage{hhline}	
\usepackage{mathrsfs}
\usepackage{graphicx}
\usepackage{dcolumn}
\usepackage{bm}
\usepackage{multirow}
\usepackage{booktabs}
\usepackage{afterpage}
\usepackage{amsmath}
\usepackage{braket}
\usepackage{ulem}

\usepackage[utf8]{inputenc}
\usepackage[T1]{fontenc}
\usepackage{mathptmx}
\usepackage{etoolbox}

\makeatletter
\def\@email#1#2{%
 \endgroup
 \patchcmd{\titleblock@produce}
  {\frontmatter@RRAPformat}
  {\frontmatter@RRAPformat{\produce@RRAP{*#1\href{mailto:#2}{#2}}}\frontmatter@RRAPformat}
  {}{}
}%
\makeatother

\title{First-principles Study of Metallic-atom Diffusion in Thermoelectric Material Mg$_3$Sb$_2$}

\author{Masayuki Ochi$^{1,2}$\thanks{ochi@phys.sci.osaka-u.ac.jp}, Kazutaka Nishiguchi$^3$, Chul-Ho Lee$^4$, and Kazuhiko Kuroki$^1$}
\inst{$^1$Department of Physics, Osaka University, Machikaneyama-cho, Toyonaka, Osaka 560-0043, Japan \\
$^2$Forefront Research Center, Osaka University, Machikaneyama-cho, Toyonaka, Osaka 560-0043, Japan \\
$^3$Graduate School of System Informatics, Kobe University, Rokkodai-cho, Nada-ku, Kobe 657-8501, Japan \\
$^4$National Institute of Advanced Industrial Science and Technology (AIST), Tsukuba, Ibaraki 305-8568, Japan}

\abst{Mg$_3$Sb$_2$ is a promising thermoelectric material that consists of nontoxic and earth-abundant elements.
We investigate metallic-atom diffusion in Mg$_3$Sb$_2$ by calculating the defect formation energy and the diffusion energy barrier for several kinds of metallic-atom impurities.
We find that early transition metals, including $4d$ elements, with a large atomic radius have a high defect formation energy, whereas Mg and late transition metals such as Ni, Cu, and Zn have relatively low formation energies as interstitial impurities.
Interstitial Ni, which is found to have a very low defect formation energy, might diffuse in the $ab$ plane at high temperatures with the energy barrier of 0.7 eV, while it seems difficult to diffuse in the $c$ direction. Interstitial Cu has a higher defect formation energy than Ni but has a low energy barrier of $\sim$0.4 eV for diffusion in the $ab$ plane.
This study will offer important knowledge for developing a thermoelectric device of Mg$_3$Sb$_2$.}

\begin{document}
\maketitle

\section{Introduction}
Thermoelectric effects, which provide a way to harvest waste heat and convert it into electricity, is one of the most promising energy technologies.
Among many thermoelectric materials, Mg$_3$Sb$_2$ has attracted attention owing to its nontoxic and earth-abundant constituent elements.
While $p$-type Mg$_3$Sb$_2$ has been originally studied~\cite{Mg3Sb2_1, Mg3Sb2_2, Mg3Sb2_3, Mg3Sb2_4, Mg3Sb2_5, Mg3Sb2_6, Mg3Sb2_7, Mg3Sb2_8, Mg3Sb2_9}, a considerably high dimensionless figure of merit $ZT$, such as $ZT=1.5$ at 716 K for a nominal composition of Mg$_{3.2}$Sb$_{1.5}$Bi$_{0.49}$Te$_{0.01}$~\cite{Tamaki_ntype}, was achieved in $n$-type samples~\cite{Pedersen_ntype, Mn_ntype, Tamaki_ntype, Zhang_ntype}. Here, $ZT=\sigma S^2 T/\kappa$ is a well-known metric for evaluating the efficiency of energy conversion, where $\sigma$, $S$, $T$, and $\kappa$ are the electrical conductivity, Seebeck coefficient, absolute temperature, and thermal conductivity, respectively.

Mg$_3$Sb$_2$ crystallizes in the trigonal $P\bar{3}m1$ space group, which can be viewed as the Mg$^{\delta +}$ cationic layers intercalated between the [Mg$_2$Sb$_2$]$^{\delta -}$ anionic layers~\cite{Mg3Sb2_review}.
While a usual Zintl-phase compound AB$_2X_2$ consists of the ionic A$^{\delta +}$ layers and the covalent [B$_2X_2$]$^{\delta -}$ layers,
it was pointed out that the chemical bonding in Mg$_3$Sb$_2$ is mostly ionic with partial covalency both for the interlayer and intralayer bonds~\cite{chemical_bond}.
Experimentally observed low thermal conductivity~\cite{thermal_cond_expt1, thermal_cond_expt2, thermal_cond_expt3, thermal_cond_expt4, thermal_cond_expt5} has prompted theoretical studies on the lattice thermal conductivity of this material~\cite{Peng_lattice_therm, Maccioni_lattice_therm}.
It was pointed out that the low thermal conductivity originates from the strong anharmonicity of the lattice~\cite{Peng_lattice_therm}, which can be related to the aforementioned peculiar bonding character~\cite{Mg3Sb2_review}.
The relationship between the low thermal conductivity and lone-pair electrons in Zintl-phase compounds is also intriguing~\cite{lone_pair1, lone_pair2}.
The valence band structure of Zintl-phase compounds has attracted much interest from the viewpoint of orbital engineering, where the energy offset between the non-degenerate $p_z$ band and the doubly-degenerate $p_{x,y}$ bands affects electron transport~\cite{orbital_engineering, elph_valley}.
It was pointed out that the high thermoelectric performance of $n$-type Mg$_3$Sb$_2$ is partially due to the multivalley structure of the conduction bands~\cite{Zhang_ntype, elband_Singh, FScomplex, Usui_zintl}.

As in many thermoelectric materials, chemical doping and substitution are key to high thermoelectric performance in Mg$_3$Sb$_2$.
It was theoretically pointed out that the negatively charged Mg vacancy ($V_{\mathrm{Mg}}^{2-}$) has a low formation energy~\cite{Tamaki_ntype,Ohno_defect}, due to which many experiments result in $p$-type behavior~\cite{Mg3Sb2_2, Mg3Sb2_3, Mg3Sb2_4, Mg3Sb2_5, Mg3Sb2_6, Mg3Sb2_7, Mg3Sb2_8, Mg3Sb2_9}, and the $n$-type doping has been difficult in Mg$_3$Sb$_2$.
Thus, in some experimental studies on $n$-type Mg$_3$Sb$_2$, excess Mg was prepared in synthesis~\cite{Tamaki_ntype}.
Chalcogen substitution for Sb was found to be effective in achieving $n$-type doping~\cite{Pedersen_ntype, Tamaki_ntype, Zhang_ntype, chalco}.
Group-3 elements such as La and Y can also introduce the $n$-type carrier, as verified in experimental~\cite{Ladoped, Ydoped, Kihou} and theoretical~\cite{Gorai_calc, Gorai_calc_group3} studies.
Chemical doping of transition metals (Mn, Fe, Co, Cu, Nb, Hf, and Ta) is also an important way to control the thermoelectric performance of Mg$_3$Sb$_2$~\cite{Mn_ntype, Mn_ntype2, Nb, Fe_Co_Hf_Ta, Cu, Gorai_calc}.

In addition to the aforementioned aspect, defect control is important also for investigating metallic-atom diffusion in a thermoelectric device.
Usually, a thermoelectric device consists of thermoelectric elements and electrode metal bonded by glue. Generally, diffusion from the electrode and glue to  thermoelectric elements can degrade the contact and/or thermoelectric performance of elements. In particular, it seriously affects the lifetime of the device. To avoid this problem, a barrier layer can be inserted between the glue and thermoelectric elements. For this purpose, the barrier layer must be formed with metals that will not penetrate to the thermoelectric elements.
However, while several studies on the defect formation energy have been reported, as mentioned in the previous paragraph, theoretical investigation along this line is missing.

In this study, we investigate the defect formation energy and the diffusion energy barrier for several metallic atoms in the Mg$_3$Sb$_2$ crystal to see whether metallic atoms can diffuse in Mg$_3$Sb$_2$.
To evaluate the diffusion energy barrier, we used the nudged elastic band (NEB) method~\cite{NEB1,NEB2}.
We find that early transition metals, including $4d$ elements, with a large atomic radius have a high defect formation energy, whereas Mg and late transition metals such as Ni, Cu, and Zn have relatively low formation energies as interstitial impurities.
Interstitial Ni, which is found to have a very low defect formation energy, might diffuse in the $ab$ plane at high temperatures with the energy barrier of 0.7 eV, while it seems difficult to diffuse in the $c$ direction. Interstitial Cu has a higher defect formation energy than Ni but has a low energy barrier of $\sim$0.4 eV for diffusion in the $ab$ plane.
Our study reveals important knowledge for developing a thermoelectric device of Mg$_3$Sb$_2$.

\section{Methods\label{sec:cal}}

We performed first-principles calculations based on the density functional theory.
We used the Perdew--Burke--Ernzerhof parameterization of the generalized gradient approximation (PBE-GGA)~\cite{PBEGGA} and the projector augmented wave (PAW) method~\cite{paw} as implemented in the Vienna {\it ab initio} simulation package~\cite{vasp1,vasp2,vasp3,vasp4}.
The core electrons in the PAW potential were [He] for Mg, [Ne] for Al, Ti, V, Cr, and Co, [Ar] for Fe, Ni, Cu, and Zn, [Ar]$3d^{10}$ for Nb and Mo, and [Kr]$4d^{10}$ for Sb.
The plane-wave cutoff energy of 650 eV for Kohn--Sham orbitals was used.
Spin-polarized calculation was performed except for pristine, i.e., nondoped, Mg$_3$Sb$_2$. We did not include the spin-orbit coupling (SOC).
We verified that, even for the $4d$ series elements Nb and Mo, the defect formation energy changes only about 10 meV by including SOC, which is negligible in this study.

First, we optimized the structural parameters, i.e., the internal coordinates and the lattice constants, for pristine Mg$_3$Sb$_2$.
Structural optimization was performed until the Hellmann--Feynman force on every atom becomes less than 0.01 eV\ \AA$^{-1}$. 
$12\times 12\times 6$ and $18\times 18\times 10$ $\bm{k}$-meshes were used for the structural optimization and density-of-states (DOS) calculation for pristine Mg$_3$Sb$_2$, respectively.

To represent a point defect, we considered a $3\times 3\times 2$ supercell of Mg$_3$Sb$_2$ containing 18Mg$_3$Sb$_2$ and introduced a single point defect by atomic doping or substitution.
For some kinds of interstitial impurity defects, we verified that the defect formation energy calculated using the $3\times 3\times 2$ supercell agrees with those calculated using the $3\times 3\times 3$ and $4\times 4\times 2$ supercells within 0.1 eV, meaning that the $3\times 3\times 2$ supercell is sufficiently large for our purpose.
For defect calculation, we used the lattice constants optimized for pristine Mg$_3$Sb$_2$ and optimized only atomic coordinates, as adopted in several defect calculations.
This treatment is verified since we are interested in the dilute limit of defect concentration.
We considered Mg, Al, Ti, V, Cr, Fe, Co, Ni, Cu, Zn, Nb, and Mo as impurity elements.
A $4\times 4\times 3$ $\bm{k}$-mesh was used for these supercell calculations.

The defect formation energy was evaluated as the following energy difference:
\begin{equation}
E_{\mathrm{form}} = E [ D ; \mathbf{N} ] - N_1N_2N_3 E_{\mathrm{P}}
 - \sum_i n_i \mu_i ,\label{eq:Ef}
\end{equation}
where $\mathbf{N}=(N_1, N_2, N_3) = (3, 3, 2)$, $E [ D ;\mathbf{N}]$, and $E_{\mathrm{P}}$ represent the supercell size, the total energy of the supercell with a defect, and that for the perfect unit cell without any defect, respectively. $n_i$ represents the number of removed (with a minus sign) or added (with a plus sign) atom $i$, the chemical potential of which is denoted as $\mu_i$. 
For example, $\sum_i n_i \mu_i  = \mu_i$ for a single interstitial impurity atom $i$ and $=\mu_j - \mu_i$ for a substitutional impurity replacing atom $i$ with atom $j$.
We considered the impurity-rich limit, i.e., $\mu_i$ is the total energy (per atom) of metal $i$.
To evaluate the total energy of the elemental crystal of atom $i$, we calculated the ferromagnetic states for Fe, Co, and Ni, the antiferromagnetic state for Cr, and the non-spin-polarized state for other elements.

\begin{figure}
\begin{center}
\includegraphics[width=7.5 cm]{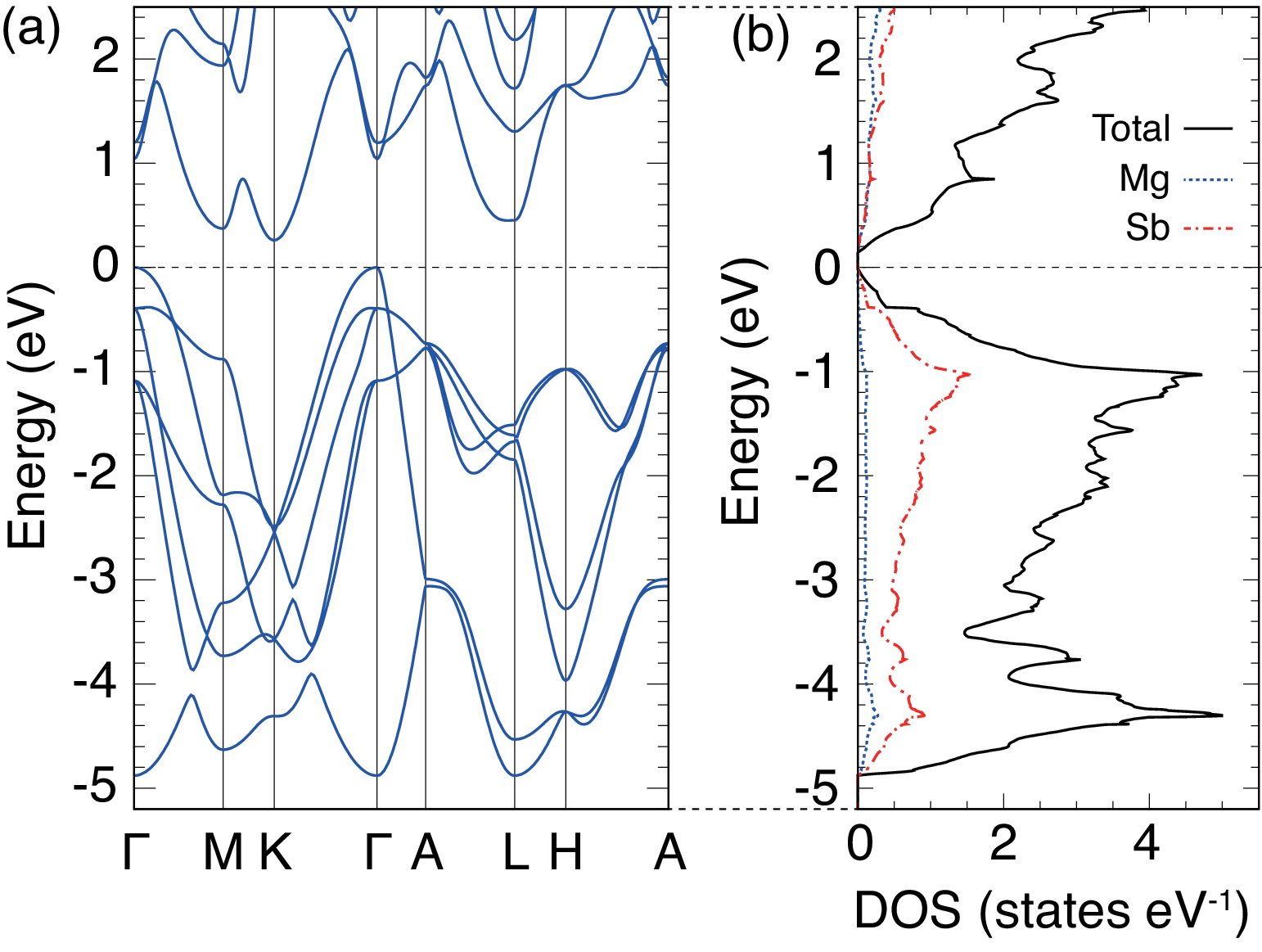}
\caption{(Color online) (a) First-principles band structure and (b) partial DOS of pristine Mg$_3$Sb$_2$.}
\label{fig:bandDOS}
\end{center}
\end{figure}

For simplicity, we did not consider several energy corrections depending on the charge of the defect and/or the Fermi level, because the band gap of Mg$_3$Sb$_2$ calculated with PBE-GGA is very small, $\sim 0.1$ eV, as shown in Figs.~\ref{fig:bandDOS}(a) and \ref{fig:bandDOS}(b), which means that the system requires a small energy for adding or removing electron carriers at least within PBE-GGA.
For a few important cases, i.e., defects with a small formation energy, we also calculated the defect formation energy using a hybrid functional offering a more accurate band gap, as presented later in this paper.

To evaluate the energy barrier for the diffusion of an impurity atom, we used the NEB method~\cite{NEB1,NEB2}.
In NEB calculations, the convergence criteria for the Hellmann-Feynman force was set to 0.025 eV\ \AA$^{-1}$.
The spring constant was set to 5 eV\ \AA$^{-2}$.
All calculations in this study were performed at zero temperature to consider a well-defined energy curve on each reaction pathway in the NEB method. 
In the NEB method, the finite-temperature effect is discussed by comparing the defect formation energy and/or the energy barrier with the temperature.
This treatment is valid as long as one considers a sufficiently lower temperature than the melting temperature.
In this study, we focus on material properties at several hundreds of Kelvin where a thermoelectric device of Mg$_3$Sb$_2$ will be used, which is sufficiently lower than the melting temperature of Mg$_3$Sb$_2$ of $\sim$ 1400--1500 K.

\section{Results and Discussion}
\subsection{Defect formation energy}

\begin{figure*}
\begin{center}
\includegraphics[width=12 cm]{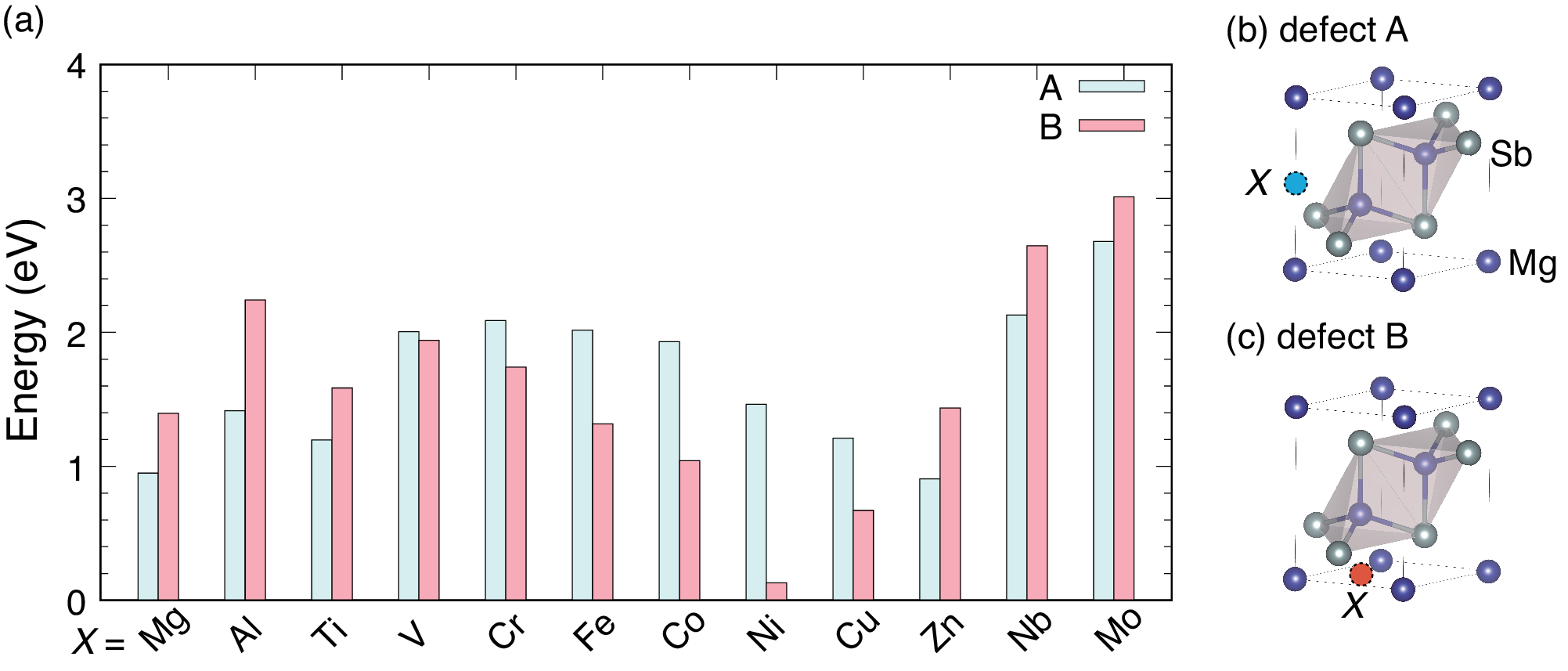}
\caption{(Color online) (a) Formation energies for defects A and B defined in panels (b) and (c), respectively. In the crystal structure, purple and gray spheres represent Mg and Sb atoms, respectively. An interstitial impurity atom $X$ is also shown. The crystal structure was depicted using the VESTA software~\cite{VESTA}.}
\label{fig:AB}
\end{center}
\end{figure*}

\begin{figure*}
\begin{center}
\includegraphics[width=14 cm]{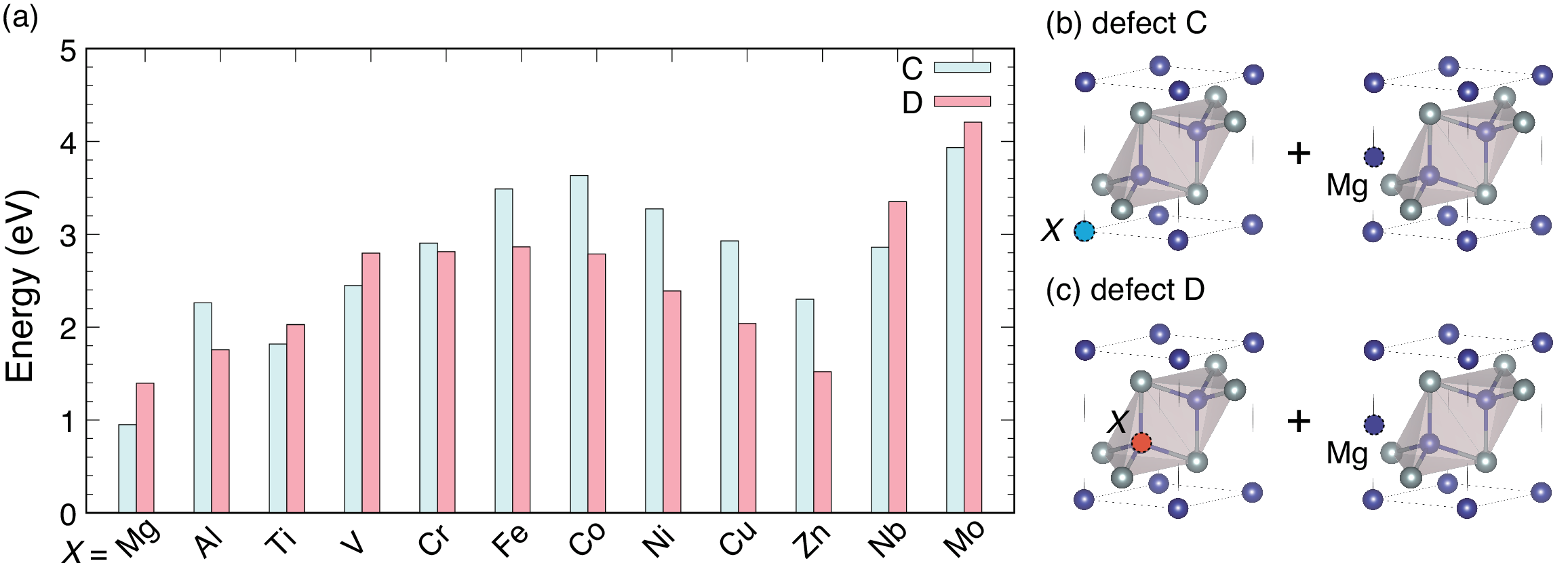}
\caption{(Color online) (a) Formation energies for defects C and D defined in panels (b) and (c), respectively. An impurity atom $X$ that substitutes Mg ($X_{\mathrm{Mg}}$) is shown in the crystal structure.}
\label{fig:CD}
\end{center}
\end{figure*}

Figure~\ref{fig:AB}(a) shows the calculated defect formation energy for two kinds of point defects shown in Figs.~\ref{fig:AB}(b) and \ref{fig:AB}(c).
Here, we considered two possible interstitial sites to place an impurity atom $X$, which we hereafter call sites A and B for defects A and B, respectively.
Interstitial sites A and B were considered, as in Refs.~\citen{Tamaki_ntype} and ~\citen{Frenkel}, respectively. 
We find that atoms with a large atomic radius, early transition metals including $4d$ elements (V, Cr, Nb, and Mo), have large formation energies, as naturally expected.
On the other hand, Mg and late transition metals such as Ni, Cu, and Zn have relatively low formation energies of $<$ 1 eV.
Defect B is preferred for $X=$ V, Cr, Fe, Co, Ni, and Cu, while other atoms prefer defect A.
The A-site preference for $X=$ Mg is consistent with the results of a previous study considering interstitial Mg placed at site A~\cite{Tamaki_ntype}.
A relatively low defect formation energy for $X=$ Cu is consistent with that in Ref.~\citen{Cu}.

As another possibility, a cationic impurity atom $X$ can substitute Mg instead of occupying an interstitial site.
To investigate this possibility, we calculated the defect formation energy for the case where $X$ substitutes Mg and the Mg atom kicked out by $X$ moves away to an interstitial site, as shown in Fig.~\ref{fig:CD}. Note that the interstitial Mg atom was assumed to occupy site A since we verified the A-site preference of interstitial Mg in Fig.~\ref{fig:AB}(a).
Figures~\ref{fig:AB}(a) and ~\ref{fig:CD}(a) show that formation energies for defects C and D are basically higher than those for defects A and B for each element, which means that $X_{\mathrm{Mg}}$ substitution is energetically unfavorable.
Therefore, we focus on defects A and B hereafter for simplicity.

Note that the defect formation energy for $X_{\mathrm{Mg}}$ becomes lower by 0.95 eV assuming that a Mg atom kicked out by an impurity atom $X$ crystallizes as Mg metal outside Mg$_3$Sb$_2$ instead of existing inside Mg$_3$Sb$_2$ as defect A, because the formation energy of defect A for $X=$ Mg is 0.95 eV, as presented in Fig.~\ref{fig:AB}(a).
In that case, the $X_{\mathrm{Mg}}$ defect is more stable than defects A and B for some elements. In fact, previous studies showed that some metal atoms substitute Mg rather than existing as interstitial impurities~\cite{Gorai_calc, Gorai_calc_group3}.
While this kind of defect should be considered when one discusses, for example, chemical doping through materials synthesis, we here considered metallic-atom penetration and diffusion during the use of the thermoelectric device where strong crystal deformation is expected to be absent. Thus, in this study, we assumed that a Mg atom driven out from its original position would exist as an interstitial impurity. 
Note also that the previous study revealed that the Mg vacancy, which introduces hole carriers, can become stable under the Mg-poor condition~\cite{Tamaki_ntype}.
If a substantial number of Mg vacancies exist in a crystal, an impurity metallic atom can conduct using vacancy sites.
However, in this study, we focus on $n$-type Mg$_3$Sb$_2$ with a Mg-rich environment, which exhibits high thermoelectric performance in experiments, and thus do not consider the metallic-atom conduction using the Mg vacancy site.

\subsection{Defect formation energy evaluated using the HSE03 hybrid functional}
Before proceeding to the next subsection, we present the defect formation energies evaluated using the HSE03 hybrid functional~\cite{HSE03_1, HSE03_2}.
Here, we used HSE03 because the band gap calculated for pristine Mg$_3$Sb$_2$ using a $9\times 9 \times 5$ ${\bm k}$-mesh with structural optimization was 0.58 eV, which is in good agreement with the experimental band gap of 0.5--0.6 eV~\cite{gap_expt}.
Since the hybrid-functional calculations are computationally demanding, we focused on the important interstitial impurities: defect A with $X=$ Zn and defect B with $X=$ Cu and Ni, which have low ($< 1$ eV) formation energies in PBE-GGA calculations, as presented in Fig.~\ref{fig:AB}.

For calculating the defect formation energy using HSE03, we should consider charged defects since the band gap is not very small compared with that evaluated by PBE-GGA.
Thus, we used the following equation for a point defect $D$ in charge state $q$ instead of Eq.~(\ref{eq:Ef})~\cite{defect1,defect2}:
\begin{equation}
E_{\mathrm{form}}[D^q ; \mathbf{N}](\Delta \epsilon_{\mathrm{F}}) = E [ D ; \mathbf{N} ] - E_{\mathrm{P} ; \mathbf{N}} 
 - \sum_i n_i \mu_i q (\epsilon_{\mathrm{VBM}} + \Delta \epsilon_{\mathrm{F}}),\label{eq:Ef_HSE}
\end{equation}
where 
$E_{\mathrm{P} ; \mathbf{N}}$ is the total energy of pristine Mg$_3$Sb$_2$ in the $N_1 \times N_2 \times N_3$ supercell,
$\epsilon_{\mathrm{VBM}}$ is the energy level of the valence-band maximum (VBM),
and $\epsilon_{\mathrm{VBM}} + \Delta \epsilon_{\mathrm{F}}$ represents the Fermi level of the system.
In addition, after calculating $E_{\mathrm{form}}[D^q ; \mathbf{N}](\Delta \epsilon_{\mathrm{F}})$ for ${\bm N}=(2, 2, 2)$ and $(3, 3, 2)$ supercells, 
we assumed that the finite-size error is proportional to $(N_1 N_2 N_3)^{-1}$~\cite{defect1}, i.e., $E_{\mathrm{form}}[D^q ; \mathbf{N}](\Delta \epsilon_{\mathrm{F}}) = c_0 + c_1(N_1 N_2 N_3)^{-1}$ holds.
Then, we obtained the extrapolated value $c_0$, which equals
\begin{equation}
E_{\mathrm{form}}[D^q]( \Delta \epsilon_{\mathrm{F}}) = \lim_{N_1,N_2,N_3\to \infty} E_{\mathrm{form}}[ D^q ; \mathbf{N}] ( \Delta \epsilon_{\mathrm{F}}).
\end{equation}
For $2\times 2\times 2$ and $3\times 3\times 2$ supercells, we used $3\times 3\times 1$ and $2\times 2\times 1$ ${\bm k}$-meshes, respectively.

\begin{figure}
\begin{center}
\includegraphics[width=4 cm]{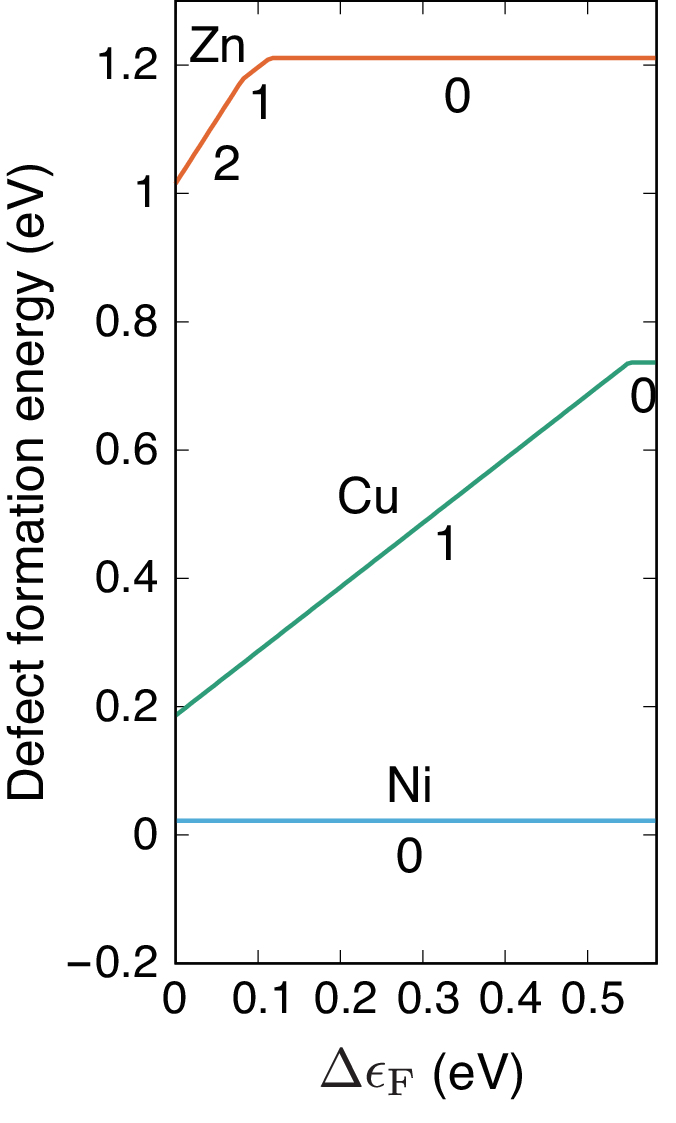}
\caption{(Color online) Formation energies of defect A for $X=$ Zn and defect B for $X=$ Cu and Ni calculated using HSE03. The value of $q$, which equals the slope of each line, is shown beside the line. The defect formation energies for the Fermi level $\Delta \epsilon_{\mathrm{F}}$ within the band gap (i.e., $\Delta \epsilon_{\mathrm{F}}$ less than the conduction band bottom) are shown.}
\label{fig:HSE}
\end{center}
\end{figure}

Figure~\ref{fig:HSE} shows the formation energies of defect A for $X=$ Zn and defect B for $X=$ Cu and Ni calculated using HSE03.
Defect formation energies at the conduction-bottom edge (i.e., those at the dilute electron-carrier doping limit, which is a target system in this study) are 0.02, 0.74, and 1.21 eV for Ni, Cu, and Zn, respectively.
These values are roughly consistent with those calculated using PBE-GGA, 0.13, 0.67, and 0.91 eV, respectively, which are shown in Fig.~\ref{fig:AB}.
Thus, we again reach the conclusion that these three types of defects, in particular, Ni and Cu, are relatively stable. 
Since NEB calculations using hybrid functionals are computationally too expensive to perform, we shall get back to PBE-GGA in the next subsection.

\subsection{Diffusion energy barrier}

\begin{figure*}
\begin{center}
\includegraphics[width=12 cm]{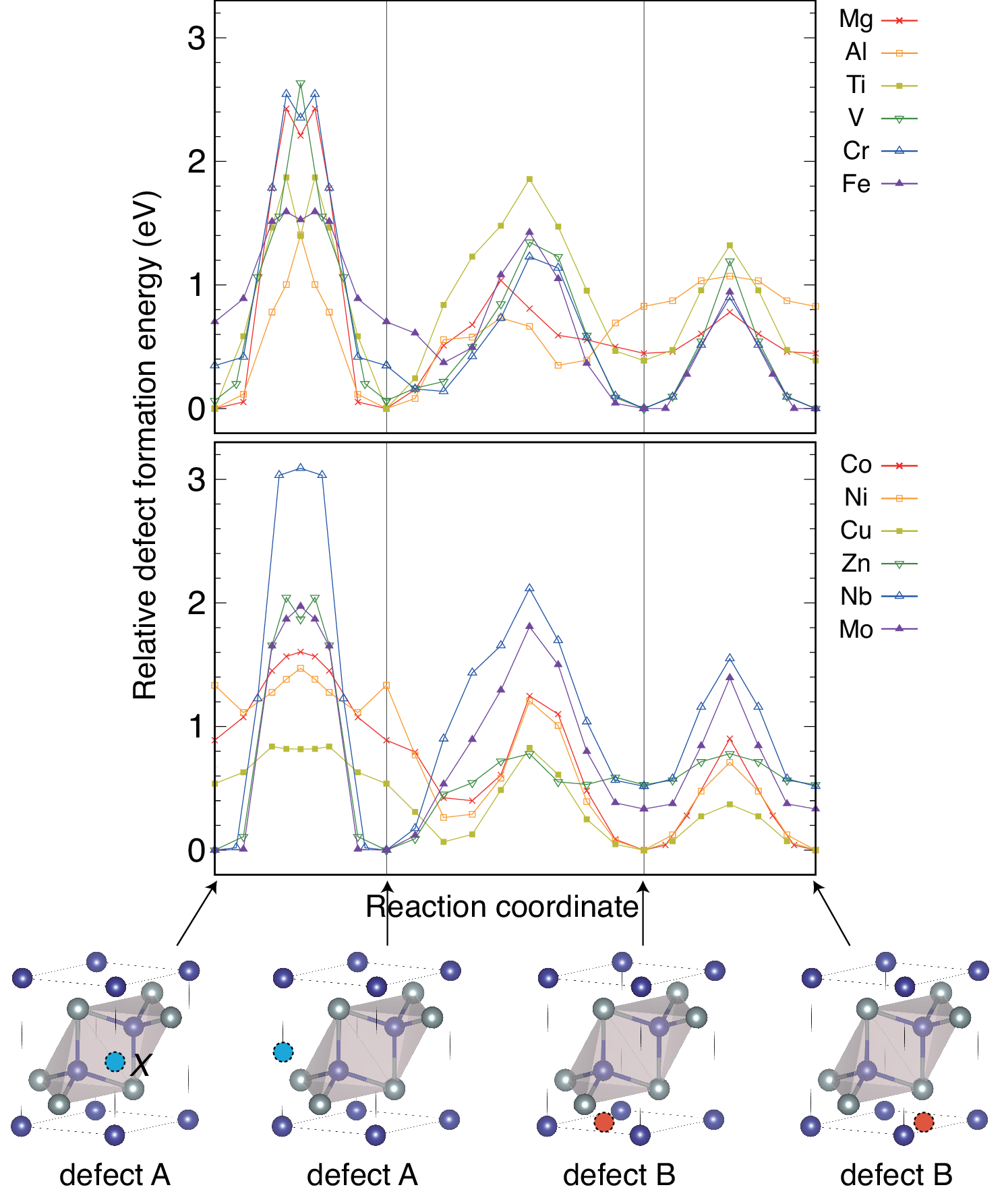}
\caption{(Color online) Energy diagram along the diffusion paths shown at the bottom for each impurity atom $X$. Defect formation energy relative to the lower value between defects A and B is shown for each impurity atom.}
\label{fig:diagram}
\end{center}
\end{figure*}

To evaluate the energy barrier for metallic-atom diffusion in Mg$_3$Sb$_2$, we focused on the three pathways depicted in Fig.~\ref{fig:diagram}; site A to site A, site A to site B, and site B to site B. While there can be other possible pathways for diffusion, we expect that comparison with respect to the same pathway among several elements will give an insight on how mobile each interstitial impurity atom is. 
Investigating more complex diffusion paths using Mg sites, such as the interstitialcy diffusion, is an important future task.
Figure~\ref{fig:diagram} shows the energy curves along the diffusion path for several kinds of impurity atoms calculated by the NEB method.
The horizontal axis represents the reaction coordinate, which is a set of atomic coordinates along the diffusion path determined by the NEB method.

Here, we focus on Cu and Ni, which have relatively low defect formation energies, as we have seen in Fig.~\ref{fig:AB}.
For Ni, which has the lowest defect formation energy at site B,
the energy barriers along the A--B and B--B paths are 1.3 and 0.7 eV, respectively.
While interstitial Ni seems to be mobile in the $ab$ plane through the B--B path at high temperatures, it seems to be difficult to travel along the $c$-axis, and the diffusion path seems to be limited in the $ab$ plane.
For Cu, our calculation shows relatively low energy barriers, 0.3, 0.4, and 0.8 eV for A--A, B--B, and A--B paths, respectively.
Since the stable interstitial site is B for Cu, the B--B path with the relatively low energy barrier of 0.4 eV might enable interstitial Cu to diffuse in the $ab$ plane.
It is interesting that the Cu diffusion is observed in some thermoelectric materials~\cite{Cu_liquid}.

\section{Summary}
To summarize, we have investigated the defect formation energy and the diffusion energy barrier for several metallic atoms in the Mg$_3$Sb$_2$ crystal. As naturally expected, early transition metals, including $4d$ elements, with a large atomic radius have a high defect formation energy whereas Mg and late transition metals such as Ni, Cu, and Zn have relatively low formation energies as interstitial impurities.
Interstitial Ni, which has been found to have a very low defect formation energy, might diffuse in the $ab$ plane at high temperatures with the energy barrier of 0.7 eV, while it seems difficult to diffuse in the $c$ direction. Interstitial Cu has a higher defect formation energy than Ni but has a low energy barrier of $\sim$0.4 eV for diffusion in the $ab$ plane.
The results of this study will offer important knowledge for developing a thermoelectric device of Mg$_3$Sb$_2$. In particular, our results confirm that we can safely use the metallic barrier layer in a thermoelectric device of Mg$_3$Sb$_2$ without considering the penetration and diffusion of metallic atoms into Mg$_3$Sb$_2$.

\acknowledgments
This study was supported by JST CREST (Grant No.~JPMJCR20Q4) and JSPS KAKENHI (Grant No.~JP22K04908), Japan.
Part of the computation was performed using the facilities of the Supercomputer Center of the Institute for Solid State Physics, The University of Tokyo, Japan.

\end{document}